\documentstyle[aaspp4]{article}

\begin{document}
\title{A major radio outburst in III~Zw~2 with an extremely inverted, millimeter-peaked spectrum}

\author{
Heino Falcke\altaffilmark{1,7}, 
Geoffrey C. Bower\altaffilmark{1,2}, 
Andrew P. Lobanov\altaffilmark{1}, 
Thomas P. Krichbaum\altaffilmark{1}, 
Alok R. Patnaik\altaffilmark{1}, 
Margo F. Aller\altaffilmark{3}, 
Hugh D. Aller\altaffilmark{3}, 
Harri Ter\"asranta\altaffilmark{4},
Melvyn C.H. Wright\altaffilmark{5},
G\"oran Sandell\altaffilmark{6}
}

\altaffiltext{1}{Max-Planck-Institut f\"ur Radioastronomie, Auf dem H\"ugel 69, D-53121 Bonn, Germany}
\altaffiltext{2}{Current address: NRAO, P.O. Box O, Socorro, NM 87801-0387}
\altaffiltext{3}{Astronomy Department, University of Michigan, Ann Arbor, MI 48109-1090}
\altaffiltext{4}{ Mets\"ahovi Radio Research Station, Metsahovintie, SF-02540 Kylm\"al\"a, Finland}
\altaffiltext{5}{Radio Astronomy Laboratory, University of California, Berkeley, CA 94720}
\altaffiltext{6}{National Radio Astronomy Observatory, P.O. Box 2, Green Bank, WV 24944}
\altaffiltext{7}{Steward Observatory, University of Arizona, Tucson, AZ 85721}

\begin{abstract}
III~Zw~2 is a spiral galaxy with an optical spectrum and faint
extended radio structure typical of a Seyfert galaxy, but also with an
extremely variable, blazar-like radio core.  We have now discovered a
new radio flare where the source has brightened more than twenty-fold
within less than two years. A broad-band radio spectrum between 1.4
and 666 GHz shows a textbook-like synchrotron spectrum peaking at 43
GHz, with a self-absorbed synchrotron spectral index $+2.5$ at
frequencies below 43 GHz and an optically thin spectral index $-0.75$
at frequencies above 43 GHz. The outburst spectrum can be well fitted
by two homogenous, spherical components with equipartition sizes of
0.1 and 0.2 pc at 43 and 15 GHz, and with magnetic fields of 0.4 and 1
Gauss.  VLBA observations at 43 GHz confirm this double structure and
these sizes. Time scale arguments suggest that the emitting regions
are shocks which are continuously accelerating particles. This could
be explained by a frustrated jet scenario with very compact hotspots.
Similar millimeter-peaked spectrum (MPS) sources could have escaped
our attention because of their low flux density at typical survey
frequencies and their strong variability.
\end{abstract}


\section{Introduction}
III~Zw~2 (PG 0007+106, Mrk 1501, z=0.089) is an active galaxy with
very unusual radio properties and extreme variability. The source was
discovered by Zwicky (1967) and initially classified as a Seyfert
galaxy (e.g. Arp 1968; Khachikian \& Weedman; Osterbrock 1977) but was
later also included in the PG quasar sample (Schmidt \& Green
1983). The galaxy was classified as a probable spiral by Hutchings et
al. (1982) which was later confirmed by fitting model isophotes to NIR
galaxy images (Taylor et al.~1996).  The bolometric, optical-to-UV,
luminosity for III~Zw~2 was estimated to be $L_{\rm UV}\sim10^{45}$
erg/sec (Falcke, Malkan, Biermann 1995; scaled to $H_0=75$ km/sec/Mpc
as used here).

The most intriguing property of III Zw 2, however, is its extreme
variability at radio and other wavelengths. Schnopper et al.~(1978)
described the onset of a huge, four year long radio outburst which
peaked around 1980. A more detailed radio lightcurve of this outburst
was presented by Aller et al.~(1985); the source showed a twenty-fold
increase in radio flux density, rising from about 100 mJy at 8 GHz to
over 2 Jy within four years. For comparison, the typical amplitude
variability in Blazars like 3C279 is usually not much larger than a
factor two or three. The source also shows optical (Lloyd 1984) and
x-ray variability (timescale less than a day; Kaastra \& Korte 1986;
Pounds 1986).

Such a pattern of variability is usually associated with BL Lac
objects and blazars, which are thought to be associated with radio
galaxies within the unified scheme (e.g.~Urry \& Padovani
1995). However, as Falcke et al.~(1995; 1996a\&b) have pointed out,
III~Zw~2 clearly does not fit into these categories. Its time averaged
radio-to-optical flux density ratio $R$ is of the order 200 (Falcke et
al.~1996a) which is significantly less than in typical blazars like
3C273 ($R\sim2000$) and other flat-spectrum radio quasars. Its
extended radio flux density from parsec to kiloparsec scales consists
of only a weak 8 mJy component at 1.4 GHz (Unger et al. 1987) and
hence cannot be associated with a typical radio galaxy. Radio spectrum
and interferometer observations also rule out a
Gigahertz-Peaked-Spectrum (GPS) or Compact-Steep-Spectrum (CSS) source
which are considered to be predecessors of giant radio galaxies (O'Dea
1998); the structure as measured with VLBI and MERLIN is point-like
(Falcke et al.~1996b; Kellermann et al.~1998).

The shape and luminosity of the extended radio emission, and the host
galaxy type are rather typical for a Seyfert galaxy and it has
therefore been suggested that III~Zw~2 could be a `weak blazar' in a
Seyfert galaxy or radio quiet quasar, i.e. an intrinsically radio
weak, yet relativistic jet pointing towards the observer. Falcke et
al.~(1995 \& 1996a) have identified a number of similar sources, named
radio-intermediate quasars (RIQs), which 
might indeed form such a class of
`weak blazars`.

To test this hypothesis we have started to watch III~Zw~2 more
carefully. In the beginning of 1998 it became clear that III~Zw~2
would show another major outburst and we launched a number of
observational projects to study the radio evolution of this outburst
and here report our first results.

\section{Observations \& Results}
\subsection{Radio lightcurves}
III~Zw~2 was re-entered into the radio-flux monitoring program at
Michigan (Aller et al.~1985) in 1997 May (observations previously
ceased at the end of 1995) and thereafter was observed every few days
at 4.8, 8, and 14.5 GHz. The 15 GHz flux density rose linearly since
then while the 8 and 4.8 GHz flux density levels remained low,
indicating an inverted spectrum. In addition, III~Zw~2 was also
monitored at 22 and 37 GHz at the Mets\"ahovi Radio Research Station
(MRRS; e.g.~Ter\"asranta, H. et al.~1992) over more than ten
years. The results are shown in Figure~1, where we have plotted the
radio light curve at 22 GHz up until 1998 September. The most recent
rise of the flux density is very obvious without any sign of a
slow-down of the flare. Its magnitude indicates the advent of a major
outburst.  A more detailed discussion of the lightcurves will be given
in a later paper.

\subsection{Radio spectrum}
In addition to the available monitoring data, we used the
Very-Large-Array (VLA), the Effelsberg 100m telescope, the BIMA array,
and the James-Clerk-Maxwell-Telescope (JCMT) to obtain a reliable and
almost simultaneous broad-band radio spectrum of III~Zw~2 extending
over two and a half decades in frequency from 1.4 to 660 GHz.

The VLA observations were made in A-configuration on 1998 May 21
simultaneously in L (1.465 GHz) and P (350 MHz) band with a total
integration time of 24 mins. 3C48 was used as the primary flux density
calibrator and III~Zw~2 was self-calibrated and mapped with AIPS. At
1.4 GHz we find flux densities of 44 mJy for the core and 8 mJy for an
extended, lobe-like component 15\arcsec{} to the SW. A heavily tapered
map reveals a possible bridge between core and lobe. The core also has
a weak extension towards the west. The core and the lobe are also seen
at 350 MHz where they are both point-like. Because the amplitudes of
the interferometer data at 350 MHz were corrupted by interference, an
absolute flux-estimate was not possible. Nevertheless, we can infer
from the data that the lobe-to-core ratio at 350 MHz is 0.5,
consistent with a flat core and a steep lobe spectrum (e.g. for a
typical, yet at this stage completely arbitrary, spectral index of
$\alpha=-1$ ($S_\nu\propto\nu^\alpha$) for the lobe we would get a
core flux density of roughly 65 mJy).

III~Zw~2 was observed with the Effelsberg 100m telescope on 1998 May
16 under very good weather conditions at 2.7, 4.9, 10.5, 14.6, 23, 32,
\& 43 GHz with cross scans (e.g., Kraus 1997). NGC~7027 was used as
primary flux density calibrator (Ott et al.~1994). If we make a linear
fit with time to the available monitoring data at 15 and 22 GHz, the
predicted flux densities for this day differ from the Effelsberg
fluxes by only 1.3 and 2.5\% respectively.  High-frequency data for
III~Zw~2 were obtained on 1998 May 20 \& 24 with BIMA at 70 through
110 GHz and with JCMT at 150, 230, 350, \& 666 GHz on 1998 May 27 \&
28 using Uranus as flux density calibrator.

The measured flux densities at the various frequencies are plotted in
Figure 2 with gray dots. The spectrum is highly inverted at cm
wavelengths, with a spectral index as high as $\alpha=+1.9\pm0.1$
between 4.8 and 10.5. The spectral turnover is around 43 GHz. Some
flattening of the spectrum can be seen towards lower frequencies.  At
frequencies above 43 GHz the spectrum becomes steep with a spectral
index around $\alpha=-0.75\pm0.15$.

\subsection{Spectral fitting}
With the available broad frequency coverage we are able to decompose
the spectrum into its various components. The spectral shape is
clearly indicative of an almost homogenous, spherical synchrotron
source which becomes self-absorbed below 43 GHz and is optically thin
at high frequencies. For such a blob where $S_{\nu}$ is the observed
flux density, $D$ is the distance from the observer, $d$ is the
diameter of the blob, and $\kappa_{\rm sync}$ and $\epsilon_{\rm
sync}$ are the synchrotron absorption and emission coefficients, the
observed spectrum will be given by a simple radiative transfer
equation (e.g. Rybicki \& Lightman 1979)

\begin{equation}
S_{\nu}={d^2\over 4D^{2}}{\epsilon_{\rm sync}\over\kappa_{\rm
sync}}\left(1+e^{-d\kappa_{\rm sync}d}\left({d\kappa_{\rm sync}\over6}-1\right)\right).
\end{equation}
For a pitch-angle averaged power-law distribution of electrons with
index $p=2.5$ (between electron Lorentz factors $\gamma_{\rm min}=1$
and $\gamma_{\rm max}=10^4$), which is in equipartition with the
magnetic field, we obtain $\kappa_{\rm sync}=2.9\cdot10^{-18}
B_0^{4.25} \nu_9^{-3.25}\,\mbox{cm}^{-1}$ and $\epsilon_{\rm
sync}=5.6\cdot10^{-20} B_0^{3.75} \nu_9^{-0.75}$, where $\nu_9$ is the
frequency in GHz and $B_0$ is the magnetic field in Gauss.

In addition to this synchrotron-spectrum we have to take into account
the quiescent spectrum of III~Zw~2 on which the outburst spectrum is
superposed and which most likely leads to the flattening of the
spectrum at lower frequencies. For simplicity we assume a flat
spectrum ($\alpha=0$). This is indicated by earlier post-flare data
and is quite typical for central cores in AGN in quiescence.  We
derive a quiescent flux density level of $\sim40$ mJy, from satisfying
the requirement that the outburst spectrum in the self-absorbed part
continues as a power-law after subtraction of the quiescent spectrum.
This quiescent flux density is consistent with the 350 MHz data and
earlier VLA flux observations at 8 GHz in 1992 October by Browne et
al.~(1998) near a flux density minimum of III~Zw~2.

The pure outburst spectrum after subtraction, shown in Fig.~2 as black
dots, can be well fitted by a two-component model. If we ignore
relativistic boosting for now (see Sec.~3.3), least-squares fitting of
the synchrotron spectrum (Fig.~2, solid line) then provides an
estimate for the diameters and magnetic fields involved. We find with
our equipartition assumption $d=120\;\mu$as \& $B=0.5$ Gauss for the
lower-frequency component and $d=50\;\mu$as \& $B=1.3$ Gauss for the
higher-frequency component. We note that inclusion of relativistic
boosting with typical Lorentz factors of a few would only change these
numbers by a factor of a few.

\subsection{VLBA observations}
Following the onset of the outburst we activated a VLBA
target-of-opportunity program to monitor the structural evolution of
the source. A first epoch observation was made on 16 February 1998 at
43 and 15 GHz. We observed with a total bandwidth of 64 MHz for a full
8 hour scan, where we spent three quarter of the available observing
time at 43 GHz and one quarter at 15 GHz.  We reduced the data using
the software packages AIPS and DIFMAP (Shepherd, Pearson, \& Taylor
1994). Fringes were detected in the III~Zw~2 data on all baselines. We
calibrated the gains using system temperature information and applied
atmospheric opacity corrections.  The amplitude gains for the stations
at Brewster, North Liberty, and Pie Town 
were ignored for the absolute amplitude calibration.

The data were then self-calibrated and mapped. For naturally weighted
maps we obtained dynamic ranges of 2000:1 at 43 GHz and 5000:1 at 15
GHz. Down to respective 3$\sigma$ levels of 1.3 mJy and 0.4 mJy no
additional components besides the central core were detected. The core
itself is resolved at 43 GHz, where the visibilities drop to one third
at 1200 M$\lambda$. It is only marginally resolved at 15 GHz. At 43
GHz we find a total flux density of 1.6$\pm$0.1 Jy consistent with the
total flux density 1.7$\pm0.1$ Jy for the source from the monitoring
data.  An elliptical fit to the UV data gives a source size (FWHM) of
90 $\mu$as$\times60\;\mu$as at PA -76$^\circ$, yielding a brightness
temperature of $3\times10^{11}$K. This elongated structure can also
been seen in the super-resolved map with a circular beam of 0.1mas
shown in Figure 3 (the nominal beam is 350$\times130\mu$as at PA
$-7^\circ$).  The fitted spherical source size at 15 GHz is
100$\pm50\mu$as for a total flux density of 0.7 Jy, giving a
brightness temperature of $5\times10^{11}$K.

The closure phases at long baselines for the 43 GHz data show a
significant deviation from zero indicating an asymmetric
structure. The non-zero closure-phases are well fit by a two-component
model of a stronger core component (1$\pm0.3$ Jy, 50$\pm15\mu$as) and
a weaker secondary component (0.6$\pm0.3$ Jy, 65$\pm30\mu$as)
separated by 65$\pm30\mu$as at PA 75$^\circ\pm10^\circ$.

\section{Summary and Discussion}
\subsection{Self-absorbed synchrotron spectra}
Because the current outburst spectrum of III~Zw~2 dominates almost the
entire radio flux density, we have the unique opportunity to study the
spectrum and evolution of such a flare over a wide frequency range
with low resolution.  In typical AGN radio flares the spectrum is much
more confused by multiple components so that VLBI techniques have to
be used to separate them, providing only very small frequency coverage
because of the frequency dependent resolution of the telescopes, and
resulting blending of the components.  

Hence, our first epoch spectrum for III~Zw~2 already provides a nice
confirmation of very basic synchrotron theory, giving a spectrum with
a textbook-like shape for the outburst component: starting in the
optically thick regime with a spectral index approaching its maximal
value of $\alpha=+2.5$ it turns over into the optically thin regime
with the typical spectral index of $\alpha=-0.75$. Moreover, the
diameters we obtain from the spectral fitting assuming equipartition 
are consistent with those we actually measure with the VLBA. 

\subsection{Variability, cooling, and particle acceleration}
Since the current outburst has begun in the beginning of 1997, the 22
GHz flux density of III~Zw~2 has risen within $\sim$600 days by at
least a factor of 20. Even compared to blazars this variability
amplitude is extreme; for example the dramatic millimeter flare in
NRAO 530 (Bower et al.~1997) had an almost ten times lower (relative)
amplitude.  From the relation for the characteristic frequency
$\nu_{\rm c}=$34 GHz $(B/$Gauss$) (\gamma_{\rm e}/100)^2$ and with our
magnetic field estimate of 1 Gauss we calculate that the typical
electron Lorentz factor at 43 GHz is $\gamma_{\rm e}=100$ and at our
highest frequency 666 GHz is $\gamma_{\rm e}=390$. The synchrotron
cooling time scales $t_{\rm sync}=7.7\cdot10^6$ sec $(B/$Gauss$)^{-2}
(\gamma_{\rm e}/100)^{-1}$ at these energies are only 50 and 14 days
respectively and hence are much shorter than the duration of the
current outburst. This directly implies that the region of enhanced
emission must also be a region of very efficient in-situ particle
acceleration.

Since the measured brightness temperature is very close to the Compton
limit (a few times $10^{11}$K, Readhead 1994) and if relativistic
boosting can be ignored (see next section) it is likely that inverse
Compton cooling of the relativistic electrons is important.  This
could lead to non-thermal optical and X-ray emission and could explain
the variability in these wavebands as seen in earlier outbursts.

\subsection{Jet nature and expansion velocity}
Another interesting question of course is what the nature of this core
actually is. As speculated in Falcke et al.~(1996a\&b) the flat
quiescent spectrum measured in the past indicates a similarity to
nuclear jets found in other AGN. This notion is strengthened by the
slightly elongated structure found at VLBI scales, the faint extension
of the core seen at a similar PA in the VLA map at 1.4 GHz, and the
presence of a lobe-like feature 15\arcsec{} to the SW. In flux density
and morphology the latter is rather typical for lobes associated with
Seyfert jets (e.g. Mrk~34 or Mrk 573; Falcke, Wilson, Simpson 1998).

Assuming that one or both components were ejected at or before the
onset of the flare, we can calculate an upper-limit to the expansion
velocity of the system. Using the duration of the outburst (600 days)
as the time scale and the component separation (0.1 pc) of the
components at 43 GHz as the size scale we calculate a characteristic
velocity, i.e. expansion velocity, of $\la0.2c$.  This could either
mean a sub-luminal jet speed or a strong shock in an initially
relativistic jet with a speed $\gg0.2c$ as proposed by Falcke et
al.~(1996a\&b). The relativistic shock scenario seems to be more
likely since we have already shown above that particle acceleration is
actually taking place. In fact, the shocks could be similar to
hotspots in radio galaxies or GPS/CSS sources. Such a scenario also
would justify neglecting relativistic boosting in Sec.~2.3.

The highly inverted spectrum of III~Zw~2 at GHz frequencies points
into a similar direction. We know that such spectral shapes at lower
frequencies mainly originate from CSS and GPS sources, where the
spectrum is dominated by jets and hotspots on scales much smaller than
the galaxy. Hence, one is tempted to call III~Zw~2 a
millimeter-peaked-spectrum (MPS) source, possibly containing
ultra-compact hotspots and operating on very short timescales compared
even to GPS sources. If other, similar MPS sources exist, they might
have easily escaped current low-frequency radio surveys and could
become important targets for mm-VLBI experiments.

\subsection{Conclusion}
III~Zw~2 remains a puzzle. Its radio core luminosity during outburst
is very high ($\sim4\cdot10^{43}$erg/sec at 350 GHz, which is
comparable to the radio luminosity of Cygnus A), yet it is apparently
located in a spiral galaxy.  Moreover, its quiescent flux density and
extended structure makes it look more like a Seyfert galaxy or
radio-quiet quasar. An initially relativistic radio jet, strongly
interacting with dense ISM at the sub-pc scale, thereby producing an
intense shock (hotspot) and slowing down the jet considerably could be
a possible explanation. 
Because of the simplicity of the source and
its very low quiescent level, monitoring of the spectral and
structural evolution of the outburst will help to test this or other
scenarios and may teach us something about the general nature of radio
flares, their expansion speeds, and their connection to particle
acceleration. To achieve this, additional monitoring of this outburst
at optical and x-ray wavelengths would be highly desirable.

\acknowledgements This research was supported by DFG grants Fa
358/1-1\&2 and by NSF Grant AST-9613998 to the University of
California, Berkeley. The BIMA array is operated by the
Berkeley-Illinois-Maryland Association under funding from the National
Science Foundation. This research has made use of data from the
University of Michigan Radio Astronomy Observatory which is supported
by the National Science Foundation and by funds from the University of
Michigan. The National Radio Astronomy Observatory is a facility of
the National Science Foundation, operated under a cooperative
agreement by Associated Universities, Inc.

\clearpage
\onecolumn

\figcaption[]{Long-term radio lightcurve of III~Zw~2 at 22 GHz  measured at Mets\"ahovi.}

\figcaption[]{Broad-band radio spectrum of III~Zw~2. Gray dots show the flux densities measured between 16 May and 28 May 1998, black dots
show the outburst spectrum with the quiescent level of 40 mJy
subtracted. The three solid gray lines represent the three fitted
components (quiescent level and two self-absorbed synchrotron
components), the black solid line is the resulting fit to the outburst
spectrum (self-absorbed synchrotron components only) and the gray
dashed line represents the fit to the total observed spectrum.}

\figcaption[]{VLBA map of III~Zw~2 at 43 GHz convolved with a beam of
100$\mu$as. The gray dots represent the two-component model fitting
the non-zero closure phases.}

\clearpage

\begin{figure*}
\plotone{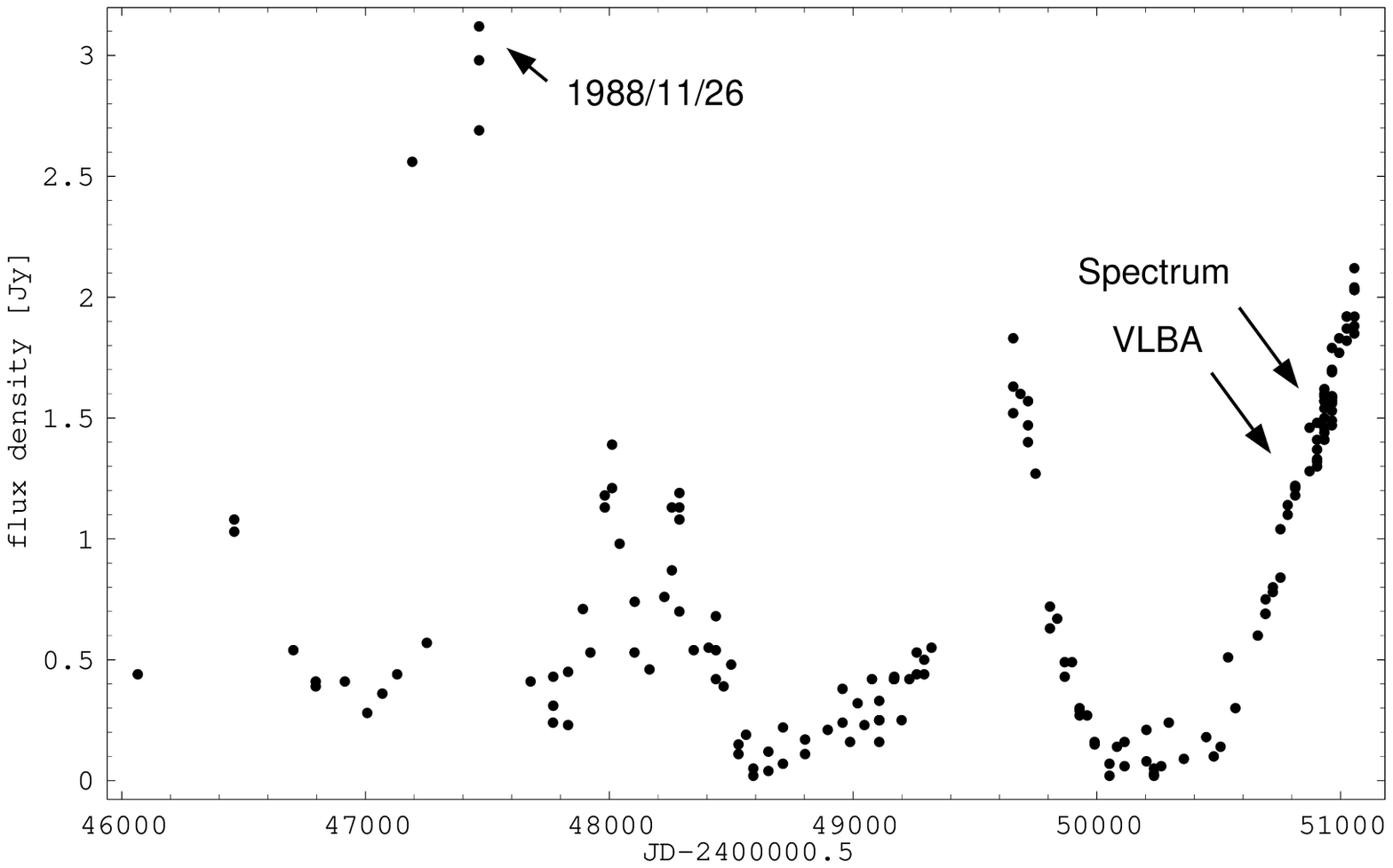}
\end{figure*}

\clearpage

\begin{figure*}
\plotone{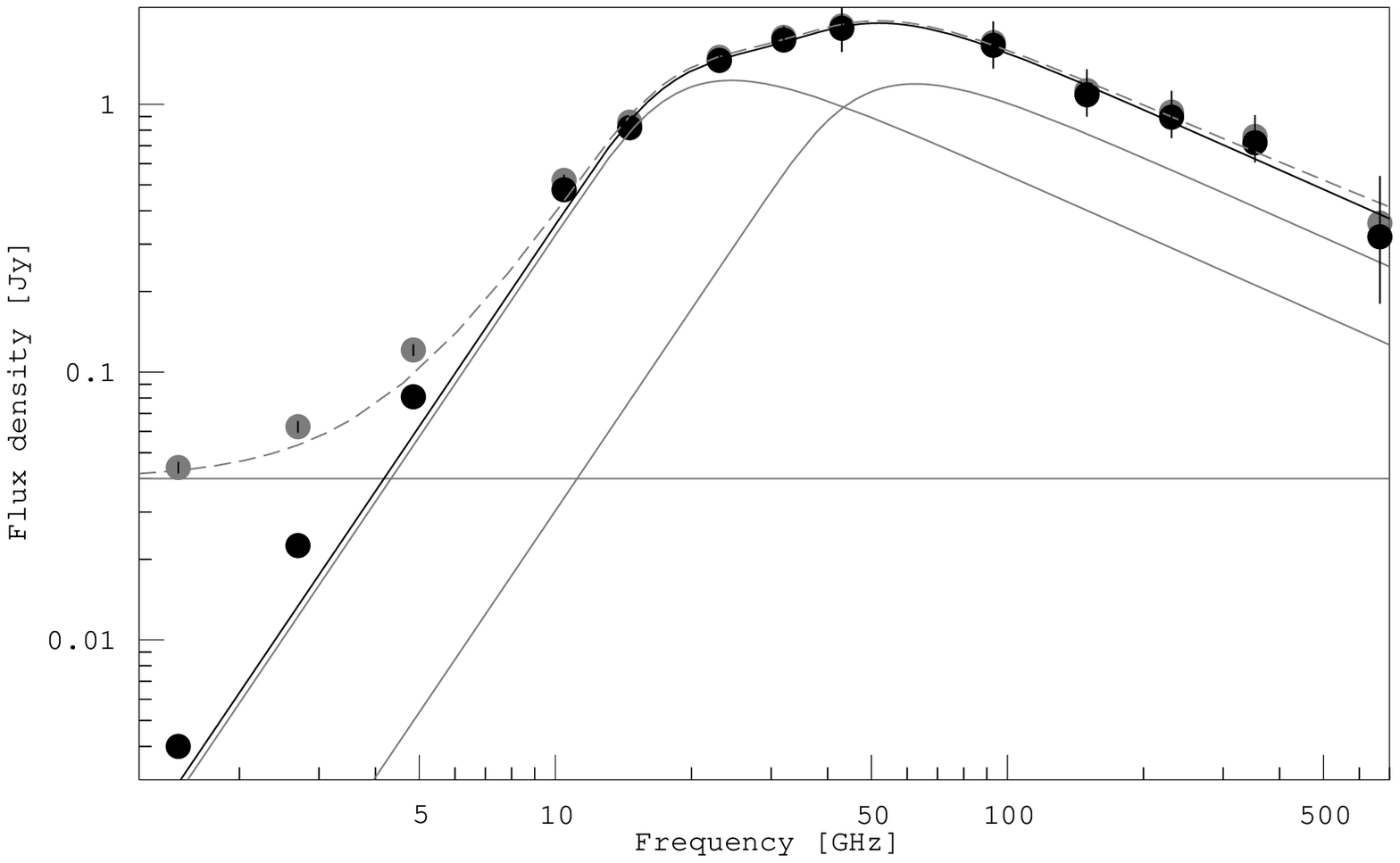}
\end{figure*}

\clearpage

\begin{figure*}
\plotone{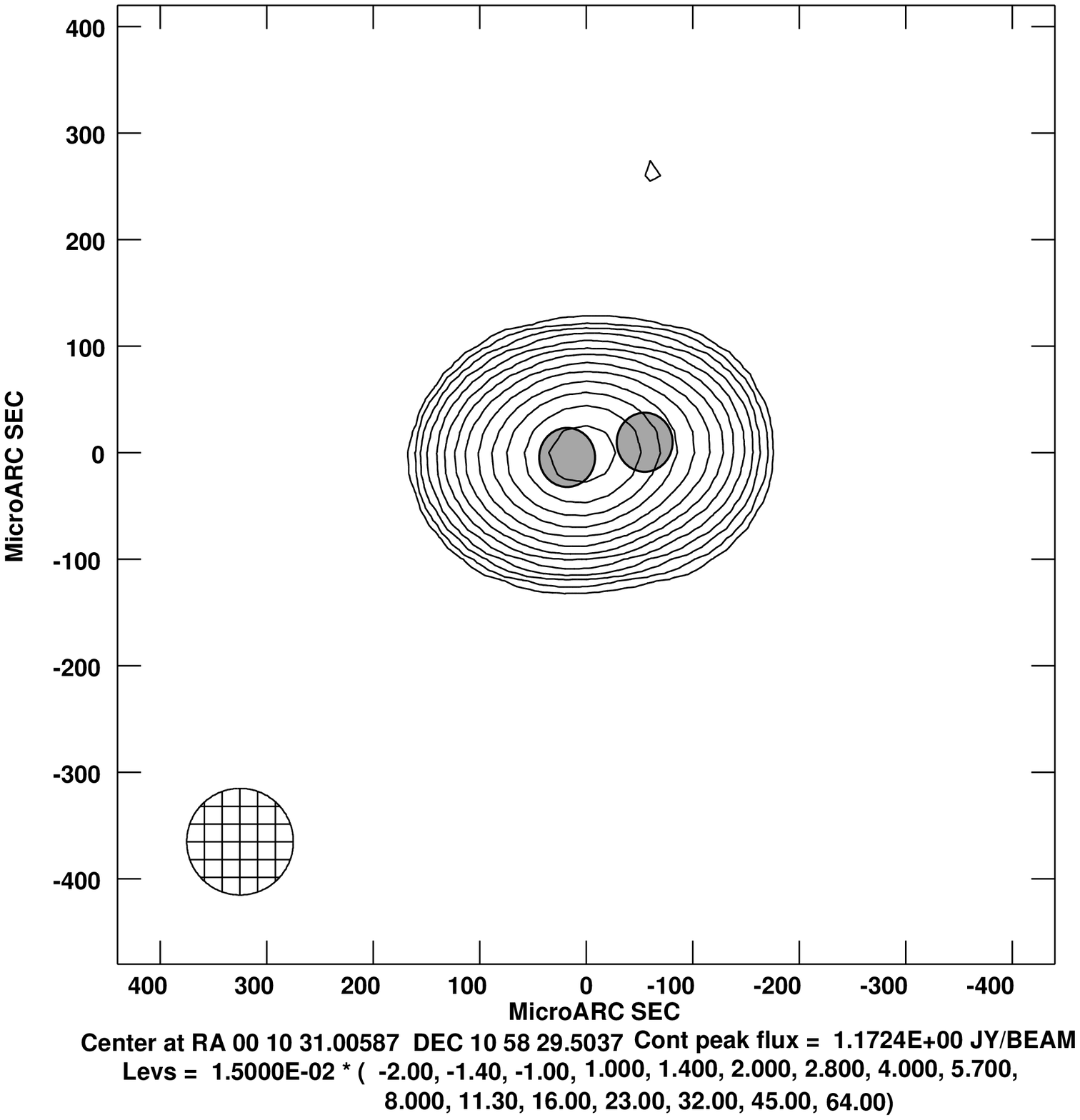}
\end{figure*}

\end{document}